# Coherent spin dynamics between electron and nucleus within a single atom

Lukas M. Veldman[1], Evert W. Stolte[1], Mark P. Canavan[1], Rik Broekhoven[1], Philip Willke[2], Laëtitia Farinacci[1], Sander Otte[1*]

[1]Department of Quantum Nanoscience, Kavli Institute of Nanoscience, Delft University of Technology, 2628 CJ Delft, The Netherlands. [2]Physikalisches Institut, Karlsruhe Institute of Technology, 76131 Karlsruhe, Germany. *e-mail: a.f.otte@tudelft.nl

**The nuclear spin, being much more isolated from the environment than its electronic counterpart, enables quantum experiments with prolonged coherence times and presents a gateway towards uncovering the intricate dynamics within an atom. These qualities have been demonstrated in a variety of nuclear spin qubit architectures[1-3], albeit with limited control over the direct environment of the nuclei. As a contrasting approach, the combination of electron spin resonance (ESR) and scanning tunnelling microscopy (STM)[4] provides a bottom-up platform to study the fundamental properties of nuclear spins of single atoms on a surface[5,6]. However, access to the time evolution of these nuclear spins, as was recently demonstrated for electron spins[7,8], remained a challenge. Here, we present an experiment resolving the nanosecond coherent dynamics of a hyperfine-driven flip-flop interaction between the spin of an individual nucleus and that of an orbiting electron. We use the unique local controllability of the magnetic field emanating from the STM probe tip[9] to bring the electron and nuclear spins in tune, as evidenced by a set of avoided level crossings in ESR-STM. Subsequently, we polarize both spins through scattering of tunnelling electrons and measure the resulting free evolution of the coupled spin system using a DC pump-probe scheme. The latter reveals a complex pattern of multiple interfering coherent oscillations, providing unique insight into the atom's hyperfine physics. The ability to trace the coherent hyperfine dynamics with atomic-scale structural control adds a new dimension to the study of on-surface spins, offering a pathway towards dynamic quantum simulation of low-dimensional magnonics.**

Nuclear spins have shown great promise as building blocks for quantum information in molecular spin qubits[2,10], NV centers[3,11], and donors in silicon[1]. They also are an excellent resource for quantum simulation[12], magnetic sensing[13,14] and spintronics[15], and are potentially scalable via engineered molecular and atomic networks[16,17]. Their key advantage arises from their longer coherence times compared to their electron spin counterpart[18], though the intricacies of the decoherence channels depend on the exact interaction with the environment and may be difficult to describe in detail. Scanning tunneling microscopy (STM) constitutes an excellent means of investigation here, since it not only permits to address individual electron spin states in electron spin resonance (ESR) experiments with sub-nanometer resolution[4,19], but also allows for an atomically precise control of their environment[20,21]. Up to now, interactions involving the nuclear spin could be measured indirectly by probing the hyperfine coupling in ESR-STM between the nucleus and the surrounding electrons[5]. In addition, the nuclear spin of individual copper atoms could be polarized via spin pumping induced by the spin-polarized tunneling current[6]. However, accessing the coherent dynamics involving the nucleus remained challenging, due to its weak coupling to the tunneling electrons.

In this work, we show the free, coherent evolution between the nuclear spin and the electron spin in a single hydrogenated titanium atom. By fine-tuning the electronic Zeeman energy using the local field of the probe tip, we identify a parameter space where electronic and nuclear spin states hybridize. In a second step, we probe the free coherent evolution of the coupled system by electric DC pump-probe experiments. Here, we reveal an emerging beating pattern, that originates from multiple quantum oscillations with different frequencies at the points of hybridization.

We use a commercial low-temperature STM equipped with high frequency cabling to send both RF signals and nanosecond DC pulses down to the tip. The sample system consists of Ti atoms deposited on bilayer MgO islands grown on Ag(100), that become hydrogenated by residual hydrogen[22,23]. For all measurements, we use spin-polarized tips that are created by picking up co-deposited Fe atoms onto the tip apex. We study individual Ti adsorbed onto the oxygen sites of MgO – well-isolated from neighboring spins using atom manipulation (see Fig. 1a) – which exhibit an effective electron spin $S = ½$ [22] with an anisotropic g-factor **g** [23]. Throughout this work, we focus on [47]Ti isotopes, which carry a nuclear spin $I = 5/2$. Along the principal axes of the crystal field, the system is described by the following Hamiltonian:

$$\widehat{\mathcal{H}} = \sum_{i=x,y,z} \left( \mu_B g_i (B_{\text{ext},i} + B_{\text{tip},i}) \hat{S}_i + A_i \hat{S}_i \hat{I}_i + Q_i \hat{I}_i^2 \right) \quad (1)$$

With the anisotropic hyperfine and quadrupole contributions, respectively, **A** = [10, 10, 130] MHz and **Q** = [1.5, 1.5, –3] MHz (see Methods)[5]. The first term describes the Zeeman energy of the electron spin with contributions from both the external $B_{\text{ext}}$ and the tip-induced magnetic field $B_{\text{tip}}$. We neglect the effect of either of these fields on the nuclear spin, since their contributions are small compared to the other terms. As shown in Fig. 1b, for small field values this Hamiltonian results in a number of avoided crossings between states with equal total angular momentum.

**State initialization via spin pumping**

We start our investigation by applying a magnetic field of 1.5 T, which is large compared to the hyperfine interaction, in order to drive ESR transitions between the individual spin states of a [47]Ti atom. Similar to measurements of Ti on a bridge binding site of MgO [24,25], we find a large anisotropy in the hyperfine coupling ranging from 10 MHz in-plane to 130 MHz out-of-plane (see Methods). Since we aim for a regime in which the hyperfine interaction competes with the Zeeman splitting of the electron, the experiments are performed with an out-of-plane magnetic field.

For certain magnetic tips with sufficient spin-polarization, we observe that the hyperfine-split ESR peaks have different intensities, which indicates a strong polarization of the nuclear spin. Such nuclear polarization has been observed for Cu atoms on MgO [6] and was modeled by taking into account inelastic spin scattering events between the tunneling electrons and electron spin. As shown in Fig. 1c, we find that the polarization is strongly dependent on the applied bias voltage while measuring at constant current. We believe that this may be due to the bias-dependent efficiency of the spin scattering channels involved, but a more complex mechanism involving the Ti orbital excitation[23] or Pauli spin blockade[26] may be at play. We find that the effective temperature of the nuclear spin population drops below 10 mK at voltages larger than 100 mV, more than two orders of magnitude lower than the actual experimental temperature of 1.5 K.

**Tuning electron-nuclear spin entanglement**

This highly efficient spin pumping allows us to overcome a major limitation: in previous ESR-STM experiments, the frequency ranges accessible for a given temperature were limited to the spin contrast set by the Boltzmann distribution. Here, owing to the nuclear polarization, we can investigate a much lower frequency regime, in which the level of entanglement between the electron and nuclear spins can be tuned. In Fig. 2a, we show the different contributions to the energy diagram of a $^{47}$Ti in a low-field regime. When the total electronic Zeeman energy – due to the external and tip magnetic field – is comparable to the hyperfine splitting, multiple avoided level crossings occur in the spectrum. The number of avoided crossings has increased compared to Fig. 1b as a result of taking into account a finite angle between the external field and the tip field. In particular, an additional avoided crossing appeared around $B_{tip}$ = 23 mT, involving a superposition of states that differ only in the electron spin, not the nuclear spin. This electron-only hybridization will be of importance below.

We identify these tuning points in our experiment by performing ESR measurements in the low field regime using an external field of merely 20 mT (Fig. 2b). Here, in order to fine-tune the coupled spin system, we vary the tip-induced magnetic field, which we adjust by the tunnel conductance $G$ of the junction. At large tip fields ($G \geq 20$ pS) multiple ESR peaks are visible in addition to several very sharp (~3 MHz) NMR type resonances around 60 MHz. The uneven splitting of the NMR type resonances is here caused by the quadrupole interaction[5]. Below $G \approx 20$ pS, the ESR and NMR transitions start to mix and overlap, accompanied by a redistribution of their intensities as shown in Fig. 2c. This is consistent with the presence of avoided level crossings between the energy levels, as expected from Fig. 2a and modelled in Figs. 2d and e. For the simulations we consider both ESR and NMR transitions with separately scaled intensities (see Methods).

**Probing coherent spin dynamics**

Having identified the appropriate tip-atom distances for inducing superposition states, we perform DC pump-probe experiments to explore the coupled spin dynamics. We use a two-pulse sequence to initialize both electron and nucleus spins to state $|\downarrow, -5/2\rangle$ via inelastic scattering[27] (See Fig. 3a and Methods). Then, after a free, varied time evolution, the final state of the system is probed by a 5 ns probe pulse.

The observed spin dynamics following this pulse sequence, shown in Fig. 3a, depend on the tip magnetic field: when the STM tip is close (i.e., at large conductance values) we observe fast, low-amplitude oscillations that become slower and stronger as the tip is retracted. This is the expected behavior when the system moves through an avoided crossing[8]. However, around $G$ = 17 pS a beating pattern appears due to interference with a second oscillation. Upon further retraction of the tip, below $G$ = 15 pS, no spin dynamics are detected anymore. Figs. 3b and c show the simulated time evolution of the $S_z$ and $I_z$ expectation values for the electron and the nuclear spin, respectively, calculated using Lindblad formalism starting from the $|\downarrow, -5/2\rangle$ state[28]. We find excellent agreement between the data and calculations, with in particular a beating pattern that arises when the electron and nucleus states are entangled. While the electron shows an interference pattern, the nuclear spin is dominated by a ~40 ns oscillation.

To understand the origin of these different oscillations, we focus on the region of interest marked in the energy spectrum of Fig. 2a: in Fig. 4a, we identify a combination of three eigenstates that form a pair of avoided level crossings, which we assign to the observed dynamics. When we tune the tip field to the second avoided level crossing at 23.7 mT, we find that the population is evenly split between state $|5\rangle$ and $|6\rangle$ corresponding to the lowest energy ESR resonance observed in Fig. 2. Here, the nucleus remains unaffected and only the electron spin forms a superposition between up and down: $|\uparrow, -5/2\rangle \pm$

$|\downarrow, -5/2\rangle$. Accordingly, we can fit in Fig. 4b a trace from the pump-probe data with a single sinusoid obtaining a frequency of roughly 75 MHz, matching the expected energy splitting between $|5\rangle$ and $|6\rangle$. The transition between states $|4\rangle$ and $|5\rangle$, on the other hand, corresponds to a flip-flop between the electron and nuclear spin. When these states hybridize they form the superposition state $|\uparrow, -5/2\rangle \pm |\downarrow, -3/2\rangle$. Thus, when the tip field is at 20.5 mT, right in middle of the two tuning points, the dynamics are expected to be a mixture between the electron oscillation discussed above and an additional flip-flop dynamic between electron and nucleus. In Fig. 4b, we observe this as a clear interference pattern in the time evolution, which we fit using three sinusoids with frequencies of 65 MHz, 20 MHz and their sum. This matches the calculated energy splitting at this tip field. Correspondingly, we can attribute the 65 MHz oscillation to a Larmor precession of the electron spin due to the in-plane component of the tip field while the 20 MHz oscillation is a flip-flop between the electron and the nucleus driven by the hyperfine interaction.

The reduced coupling of the nuclear spin to the environment is expected to result in an enhanced coherence time when the nuclear spin is involved, compared to the dynamics of only the electron spin[14,29,30]. Indeed, the data shown in Fig. 4b still shows observable dynamics up to 120 ns, whereas the oscillation in Fig. 4c has decayed already after 80 ns. We point out, however, that the increased coherence time may also in part result from a decreased sensitivity to magnetic tip field noise since the energy levels in Fig. 4a diverge less at 21 mT than at 23 mT, akin to a clock-transition[31,32].

**Conclusion**
Developing single atom quantum information processing requires thorough understanding of the underlying electron and nuclear spin dynamics. This demands initialization, tuning and readout tailored on the atomic length scale. Using pump-probe spectroscopy, we revealed the collective coherent dynamics of the internal spin dynamics inside a single atom. The magnetized STM tip functioned in this work as a control knob to locally tune the nature of these dynamics. This technique has the potential to be extended to any on-surface atomic or molecular spin system, yielding a great variety of phenomena to explore in the future. Moreover, the prospect of STM for engineering bottom-up atomic designer assemblies can provide an integral atomic-scale understanding into the fundamentals of complex coherent spin dynamics.

**References**
1   Pla, J. J. *et al.* High-fidelity readout and control of a nuclear spin qubit in silicon. *Nature* **496**, 334-338, (2013).
2   Vincent, R., Klyatskaya, S., Ruben, M., Wernsdorfer, W. & Balestro, F. Electronic read-out of a single nuclear spin using a molecular spin transistor. *Nature* **488**, 357-360, (2012).
3   Neumann, P. *et al.* Single-Shot Readout of a Single Nuclear Spin. *Science* **329**, 542-544, (2010).
4   Baumann, S. *et al.* Electron paramagnetic resonance of individual atoms on a surface. *Science* **350**, 417-420, (2015).
5   Willke, P. *et al.* Hyperfine interaction of individual atoms on a surface. *Science* **362**, 336-339, (2018).
6   Yang, K. *et al.* Electrically controlled nuclear polarization of individual atoms. *Nat Nanotechnol* **13**, 1120-1125, (2018).
7   Yang, K. *et al.* Coherent spin manipulation of individual atoms on a surface. *Science* **366**, 509-512, (2019).
8   Veldman, L. M. *et al.* Free coherent evolution of a coupled atomic spin system initialized by electron scattering. *Science* **372**, 964-968, (2021).

9   Yang, K. *et al.* Tuning the Exchange Bias on a Single Atom from 1 mT to 10 T. *Phys Rev Lett* **122**, 227203, (2019).
10  Thiele, S. *et al.* Electrically driven nuclear spin resonance in single-molecule magnets. *Science* **344**, 1135-1138, (2014).
11  Fuchs, G. D., Burkard, G., Klimov, P. V. & Awschalom, D. D. A quantum memory intrinsic to single nitrogen-vacancy centres in diamond. *Nat Phys* **7**, 789-793, (2011).
12  Randall, J. *et al.* Many-body-localized discrete time crystal with a programmable spin-based quantum simulator. *Science* **374**, 1474-1478, (2021).
13  Maze, J. R. *et al.* Nanoscale magnetic sensing with an individual electronic spin in diamond. *Nature* **455**, 644-647, (2008).
14  Zaiser, S. *et al.* Enhancing quantum sensing sensitivity by a quantum memory. *Nat Commun* **7**, 12279, (2016).
15  Zutic, I., Fabian, J. & Das Sarma, S. Spintronics: Fundamentals and applications. *Rev Mod Phys* **76**, 323-410, (2004).
16  Wernsdorfer, W. & Ruben, M. Synthetic Hilbert Space Engineering of Molecular Qudits: Isotopologue Chemistry. *Adv Mater* **31**, 1806687, (2019).
17  Kane, B. E. A silicon-based nuclear spin quantum computer. *Nature* **393**, 133-137, (1998).
18  Yang, S. *et al.* High-fidelity transfer and storage of photon states in a single nuclear spin. *Nat Photonics* **10**, 507-511, (2016).
19  Willke, P., Yang, K., Bae, Y., Heinrich, A. J. & Lutz, C. P. Magnetic resonance imaging of single atoms on a surface. *Nat Phys* **15**, 1005-1010, (2019).
20  Khajetoorians, A. A., Wiebe, J., Chilian, B. & Wiesendanger, R. Realizing All-Spin-Based Logic Operations Atom by Atom. *Science* **332**, 1062-1064, (2011).
21  Wang, Y. *et al.* Universal quantum control of an atomic spin qubit on a surface. *Npj Quantum Inform* **9**, 48, (2023).
22  Yang, K. *et al.* Engineering the Eigenstates of Coupled Spin-1/2 Atoms on a Surface. *Phys Rev Lett* **119**, 227206, (2017).
23  Steinbrecher, M. *et al.* Quantifying the interplay between fine structure and geometry of an individual molecule on a surface. *Phys Rev B* **103**, 155405, (2021).
24  Farinacci, L., Veldman, L. M., Willke, P. & Otte, S. Experimental Determination of a Single Atom Ground State Orbital through Hyperfine Anisotropy. *Nano Lett* **22**, 8470-8474, (2022).
25  Kim, J. *et al.* Anisotropic Hyperfine Interaction of Surface-Adsorbed Single Atoms. *Nano Lett* **22**, 9766-9772, (2022).
26  McMillan, S. R., Harmon, N. J. & Flatte, M. E. Image of Dynamic Local Exchange Interactions in the dc Magnetoresistance of Spin-Polarized Current through a Dopant. *Phys Rev Lett* **125**, 257203, (2020).
27  Natterer, F. D. Waveform-sequencing for scanning tunneling microscopy based pump-probe spectroscopy and pulsed-ESR. *Methodsx* **6**, 1279-1285, (2019).
28  Johansson, J. R., Nation, P. D. & Nori, F. QuTiP: An open-source Python framework for the dynamics of open quantum systems. *Comput Phys Commun* **183**, 1760-1772, (2012).
29  Degen, C. L., Reinhard, F. & Cappellaro, P. Quantum sensing. *Rev Mod Phys* **89**, 035002, (2017).
30  Savytskyy, R. *et al.* An electrically driven single-atom "flip-flop" qubit. *Sci Adv* **9**, eadd9408, (2023).
31  Bae, Y. *et al.* Enhanced quantum coherence in exchange coupled spins via singlet-triplet transitions. *Sci Adv* **4**, eaau4159, (2018).
32  Shiddiq, M. *et al.* Enhancing coherence in molecular spin qubits via atomic clock transitions. *Nature* **531**, 348-351, (2016).

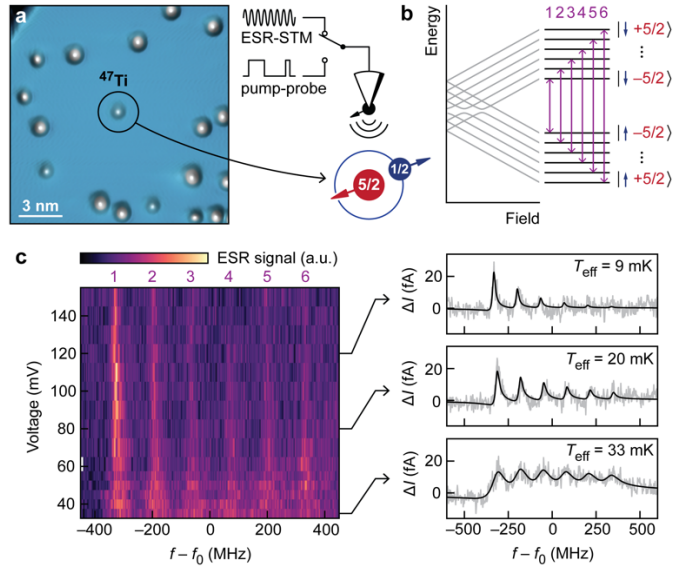

**Fig. 1 | Single atom nuclear polarization. a**, STM topography of the single $^{47}$TiH studied in this work. A schematic drawing shows the magnetic STM tip above the electron spin (blue) and nuclear spin (red) of the single atom. **b**, Energy diagram of the spin states of a single $^{47}$TiH. In the high field regime, the eigenstates resemble Zeeman product states. ESR transitions (purple arrows) can be driven between states with equal nuclear spin. **c**, ESR-STM measurements at different applied DC bias ($T$ = 1.5 K, $B_{ext}$ = 1.5 T, $V_{RF}$ = 25 mV, $I_{set}$ = 2.5 pA, $f_0$ = 11.5 – 12.56 GHz). Line traces at 35 mV, 80 mV and 120 mV are shown with fits using six Fano lineshapes scaled by the Boltzmann factor in order to extract an effective temperature.

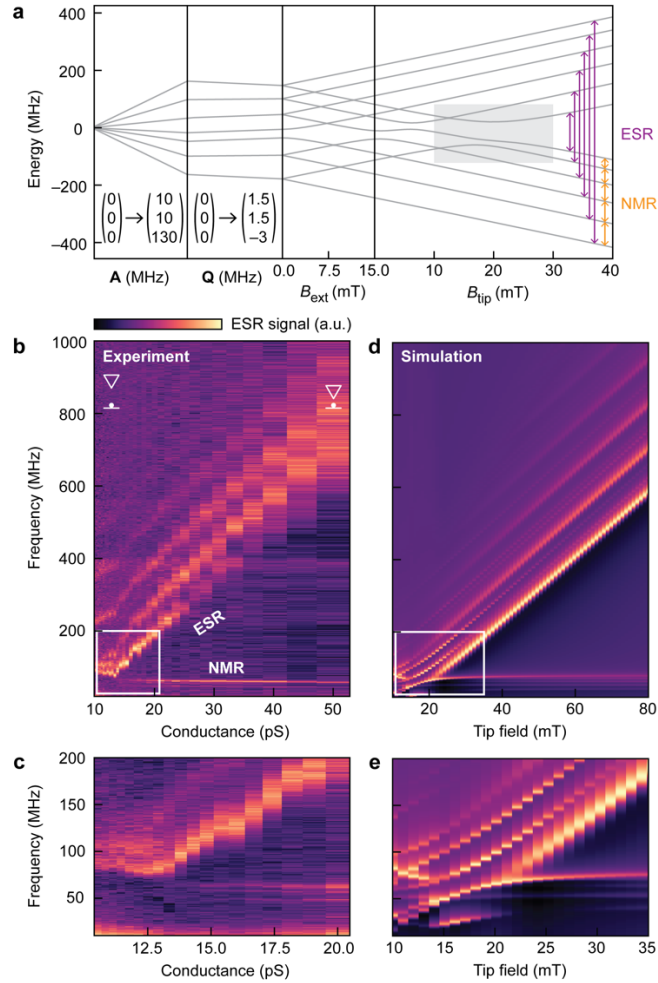

**Fig. 2 | ESR and NMR-type measurements in the low-field regime. a**, Energy diagram of the atomic eigenstates as a function of hyperfine coupling, quadrupole moment, external and tip-induced magnetic field. **b**, ESR-STM measurements ($T$ = 400 mK, $B_{ext}$ = 20 mT, $V_{RF}$ = 40 mV, $I_{set}$ = 2 pA) as function of tip-sample conductance. As indicated with icons, tip-sample separation decreases towards the right side of the plot, corresponding to an increasing tip-induced magnetic field. Both ESR and NMR-type transitions are observed and indicated as such. The bottom close-up is a separate dataset showing the splitting of the NMR transitions and a curve upwards of the bottom ESR transition signaling the avoided level crossing. **c**, Zoom-in to marked region of (b). **d**, Simulations of the ESR-STM measurements (see Methods), where we consider the tip field to make an 8° angle with the out-of-plane external field. **e**, Zoom-in to marked region of (d).

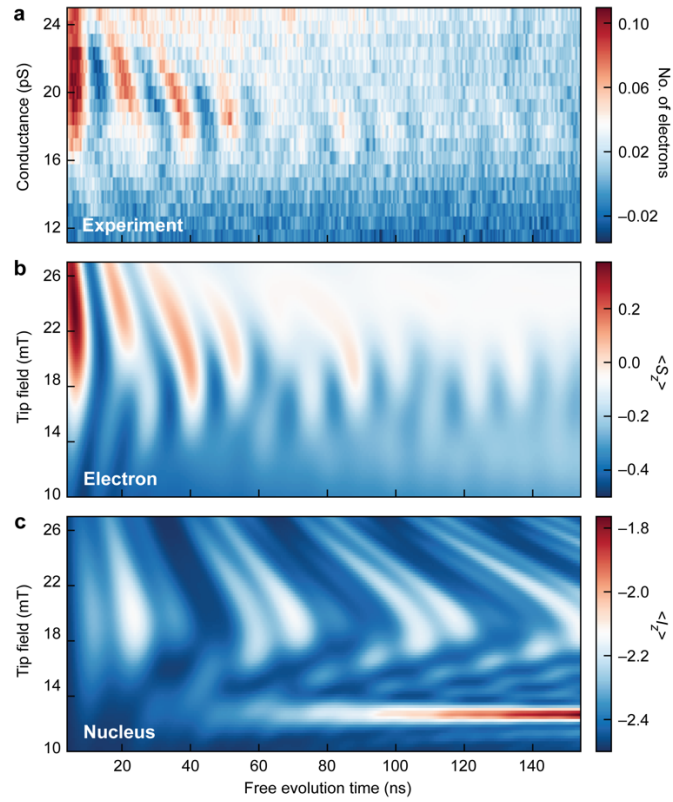

**Fig. 3 | Free evolution measurements and Lindblad simulations. a**, Pump-probe data for different tip-atom distances set by the junction conductance ($V_{set}$ = 130 mV, $T$ = 400 mK, $B$ = 15 mT, for details on the pulse scheme, see Methods). **b**, Lindblad simulation of the free time evolution of the electron spin when initialized to $|\downarrow, -5/2\rangle$. **c**, Corresponding Lindblad simulation of the free time evolution of the nuclear spin. The calculations also show the onset of an additional oscillation in the nuclear spin at around 13 mT. However, since the period is an order of magnitude longer than the coherence time of the electron spin, it is not visible in our measurements.

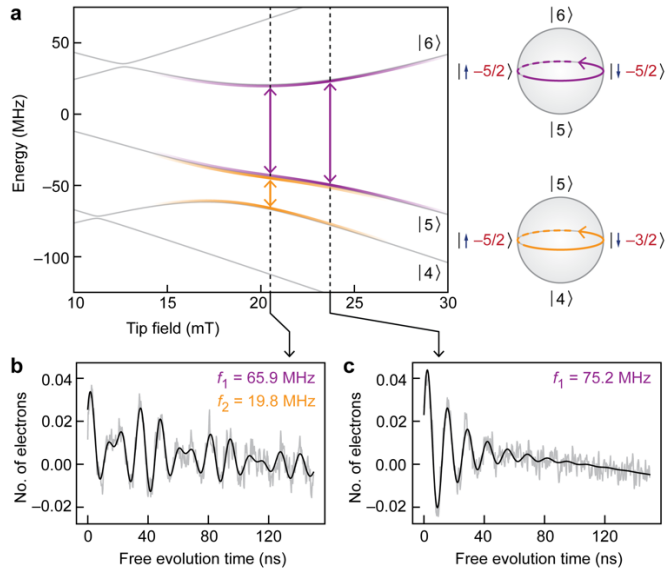

**Fig. 4 | Origin of the beating pattern. a**, Zoom-in on the relevant avoided level crossings of Fig. 2a. The Bloch spheres illustrate the dominating dynamics arising from the superpositions of the corresponding states. **b**, Fit to a line trace from the pump-probe data from Fig. 3a showing a beating pattern. The pattern arises from two dominating frequencies and their sum frequency. **c**, Fit to a line trace from the pump-probe data from Fig. 3a showing a single frequency oscillation.

## Methods

### Sample preparation
The sample was prepared in-situ in a separate vacuum chamber with a residual pressure of $10^{-10}$ mbar. The surface of the Ag(100) crystal was cleaned by two cycles of $Ar^+$ sputtering and annealing to 650°C. Subsequently, bilayer MgO islands were grown by evaporating Mg in an oxygen environment of $10^{-6}$ mbar while the sample was annealed to ~500°C. We deposited Fe and Ti atoms on the surface by e-beam evaporation. Magnetic tips were created by picking up 3 to 15 Fe atoms with a tungsten wire STM tip.

### Equipment
The measurements were conducted using a commercial Unisoku USM-1300 low-temperature STM. All DC bias values reported are with respect to the sample. For ESR measurements, we used a Rohde & Schwarz SMB100A to generate radiofrequency voltage signals and a Tektronix PSPL5542 bias-tee to combine the RF with the DC bias before sending the combined output to the STM tip. The RF signal was chopped at 271 Hz and we used a Stanford Research SR830 lock-in amplifier for detection. The ESR measurements were performed in constant-current mode. A wiring schematic of the components used during ESR measurements is shown in Extended Data Fig. 1a.

For pump-probe measurements, we used a Keysight AWG M8195a to generate the DC pulses. For details on the pulse trains, see Supplementary Section 4. The DC bias was sent through the DC input of the Tektronix PSPL5542 bias-tee and combined with the pulse train via a Mini-Circuits ZFRSC-183-S+ splitter before being sent to the STM tip. A wiring schematic of the components used during ESR measurements is shown in Extended Data Fig. 1b.

### Hyperfine anisotropy measurements
In order to correctly calculate the eigenenergies of the Hamiltonian presented in Equation (1), we need to determine the hyperfine vector **A**. Due to the $C_{4v}$ symmetry of the crystal structure of the oxygen binding site, the two in-plane directions $A_x$ and $A_y$ are equal. Therefore, we only need two measurements to characterize **A** fully: the hyperfine splitting in-plane and out-of-plane. Consequently, we sweep the angle of the external field from in-plane to out-of-plane while measuring the magnitude of the hyperfine splitting in a $^{47}$Ti atom. Results are shown in Extended Data Fig. 2: we find an in-plane splitting of 10 MHz, in accordance with previous measurements[5] and an out-of-plane splitting of 130 MHz.

### Simulations of ESR results
ESR simulations were done by generating Fano functions for each possible spin transition between two eigenstates. The amplitude $I_{nm}$ of a transition between states $|m\rangle$ and $|n\rangle$ was calculated by:

$$I_{nm} \propto |V_{nm}|^2 \Delta P_{nm} \{\mathbf{B}_{\text{tip}} \cdot (\langle n|\hat{\mathbf{S}}|n\rangle - \langle m|\hat{\mathbf{S}}|m\rangle)\} \qquad (2)$$

This amplitude consists of three components. The part between curly brackets is the readout strength, which scales with the difference in expectation value of each spin between the two states, projected onto the measurement axis. The first part corresponds to the driving amplitude between the two states. To achieve a result that matches the experiment we distinguish here between ESR and NMR type transitions and add them with separate scaling factors. The two different driving terms are modelled by:

$$\begin{aligned} V_{nm}^{\text{ESR}} &= \langle n|\mathbf{B}_{\text{tip}} \cdot \hat{\mathbf{S}}|m\rangle \\ V_{nm}^{\text{NMR}} &= \langle n|\mathbf{B}_{\text{tip}} \cdot \hat{\mathbf{I}}|m\rangle \end{aligned} \qquad (3)$$

We found the simulations to match the experimental results best when using a ratio $I_{nm}^{\text{ESR}}:I_{nm}^{\text{NMR}} = 1:100$. This difference in driving amplitudes may be explained by the nuclear spin having significant longer relaxation and coherence times than the electron spin.

The population difference $\Delta P_{nm}$ is calculated by finding the steady state solution to the Bloch-Redfield equations for the coupled spin system connected to a spin-polarized bath (tip, t) and a spin-averaged bath (sample, s). In a general form this equation is written as:

$$\frac{d\rho_{nm}(t)}{dt} = -i\omega\rho_{nm}(t) + \sum_{kl} R_{nmkl}(t)\rho_{kl}(t) \tag{4}$$

Here, the first term is the unitary von Neumann evolution of the density matrix $\hat{\rho}$ and the second term describes the influence of the environment via the Redfield tensor element $R_{nmkl}$. When a bias is applied between the two baths the resulting tunnelling electrons scatter with the atomic electron spin via Kondo interaction:

$$\hat{\mathcal{H}}_{\text{Kondo}} = \sum_{i=t,s} J\hat{\mathbf{S}} \cdot \hat{\mathbf{s}}_i \tag{5}$$

Where $\hat{\mathbf{S}}$ is the surface electron spin and $\hat{\mathbf{s}}_i$ the spin density of the tip and sample bath and $J$ the coupling between the bath and surface spin. The relevant Redfield tensor that gives rise to the spin pumping of the atomic electron spin due to the spin-polarized tunnelling electrons is given by:

$$R_{nmkl} = \frac{1}{\hbar^2}J_tJ_s \sum_{\alpha,\beta=x,y,z} \begin{pmatrix} -\delta_{ml}\sum_p \langle n|\hat{S}_\alpha|p\rangle\langle p|\hat{S}_\beta|k\rangle g_{t\to s}^{\alpha\beta}(\omega_{pk}) \\ -\delta_{nk}\sum_p \langle l|\hat{S}_\alpha|p\rangle\langle p|\hat{S}_\beta|m\rangle \left(g_{t\to s}^{\alpha\beta}(\omega_{pl})\right)^* \\ +\langle n|\hat{S}_\beta|k\rangle\langle l|\hat{S}_\alpha|m\rangle \left(g_{t\to s}^{\alpha\beta}(\omega_{nk}) + \left(g_{t\to s}^{\alpha\beta}(\omega_{ml})\right)^*\right) \end{pmatrix} \tag{6}$$

Here, the $\omega$ are the energy differences between different eigenstates. The correlation function between tip and sample baths is given by:

$$g_{t\to s}^{\alpha\beta}(\omega_{nm}) = \hbar\pi\varrho^t\varrho^s \sum_{\sigma,\sigma',\sigma''} \frac{\tau_{\sigma''\sigma}^\alpha \tau_{\sigma\sigma'}^\beta}{4} \eta_{\sigma''\sigma'}^t \eta_{\sigma\sigma}^s \iint f(\varepsilon_t)f(1-f(\varepsilon_s))\delta(\varepsilon_t - \varepsilon_s - \varepsilon_{nm} - V)d\varepsilon_t d\varepsilon_s \tag{7}$$

Where $\varrho^t$ and $\varrho^s$ are the densities of states of the tip and sample baths, $\hat{\eta}^s$ is the spin-averaged density matrix of the sample bath and $\hat{\eta}^t = \frac{1}{2}(\hat{\mathbb{I}} + \mathbf{n}\cdot\hat{\boldsymbol{\tau}})$ is the spin-polarized density matrix of the tip bath with polarization vector $\mathbf{n}$, in which $\hat{\boldsymbol{\tau}} = (\hat{\tau}^x, \hat{\tau}^y, \hat{\tau}^z)$ is a vector composed of Pauli matrices. The spin polarized electrons tunnelling between tip and sample result in an effective spin pumping of the electron spin which also affects the nucleus through the hyperfine coupling. We use the following coupling strengths between atomic electron spin and the tip and sample baths: $\varrho^t J_t$ = 0.011 and $\varrho^s J_s$ = 0.05. Coupling to each bath individually is also taken into account and leads to relaxation processes due to spin scattering.

**Technical details on pump-probe measurements**

The pump-probe measurements were performed in constant-height mode after tracking the center of the atom for ~1 hour and while linearly compensating for drift due to piezo creep. For the pump-probe data shown in the main text the tip was placed ~300 fm away from the center in order to minimize the in-plane field component of the tip.

To pump both electron and nuclear spins we start the pulse train with a 400 ns positive bias pulse. We then add an extra 5 ns negative pump pulse that we empirically found to increase the signal intensity. We hypothesize that the addition of this pulse brings the atomic electron spin in a superposition state close to $|\downarrow, -5/2\rangle$ which maximizes the amplitude of the resulting spin dynamics. In Extended Data Fig. 3 pulse sequences with and without this extra pulse are compared. To increase the signal detected by the lock-in amplifier we switch the polarity of the 5 ns probe pulse in the B-cycle[27]. All pulses went through a 5 GHz low pass filter before being sent to the STM to minimize artefacts. The pulses were calibrated to have an amplitude of 110 mV at the tip sample junction.

**Lindblad simulations**

To simulate the expected time dynamics of the coupled electron-nuclear spin system we solve the Lindblad equations for the Hamiltonian presented in the main text. Since there are no electrons tunnelling during the free evolution time, it is not necessary to use the Bloch-Redfield model that takes into account the spin polarisation of the tip bath. We only calculate the unitary von Neumann evolution and add some effective relaxation and decoherence processes in the form of Lindblad collapse operators that interact with the electron spin. In the general form the Lindblad equation is written as:

$$\frac{d\hat{\rho}(t)}{dt} = -\frac{i}{\hbar}[\hat{\mathcal{H}}, \hat{\rho}(t)] + \sum_n \frac{1}{2}\left(2\hat{L}_n \hat{\rho}(t)\hat{L}_n^\dagger - \hat{\rho}(t)\hat{L}_n^\dagger \hat{L}_n - \hat{L}_n^\dagger \hat{L}_n \hat{\rho}(t)\right) \quad (8)$$

Where the Lindblad collapse operators are given by $\hat{L}_n = \sqrt{\gamma_n}\hat{A}_n$ with $\hat{A}_n$ the operators coupling the spin to the environment and $\gamma_n$ the rates at which these processes happen. For our simulations we use the following specific Lindblad operators to simulate effective relaxation and decoherence processes occurring on the electron spin due to coupling with the tip and sample baths.

$$\hat{L}_- = \frac{1}{\sqrt{T_1}}\frac{1}{\sqrt{1 + e^{-\hbar\omega/k_\text{B}T}}}\hat{S}_-$$
$$\hat{L}_+ = \frac{1}{\sqrt{T_1}}\frac{1}{\sqrt{1 + e^{\hbar\omega/k_\text{B}T}}}\hat{S}_+ \quad (9)$$
$$\hat{L}_z = \sqrt{\frac{2T_1 - T_2}{4T_1 T_2}}\hat{S}_z$$

We use effective relaxation and decoherence times $T_1$, $T_2$ = 300 ns and a temperature $T$ = 400 mK. We separately take into account decoherence from the mechanical vibration of the tip, leading to magnetic field noise, by applying a 2 mT Gaussian filter over the magnetic field axis. These Lindblad calculations were done using a python package called QuTiP[28].

**Tip dependence**

To show the influence of the microtip, we perform ESR-STM measurements with a different microtip on the same $^{47}$Ti atom bound on the oxygen site. The results are shown in Extended Data Fig. 4 for two different $B_{ext}$ values. Compared to the data taken with the tip of the main text (Fig. 2) we find a number of differences. First, at both external field values, the lowest two ESR resonances are closer together at large conductance values and are split apart when the tip is retracted ($G$ < 40 pS). Second, the splitting between NMR resonances is so small that they are observed as a single peak that shifts in energy with applied tip field. We get a good agreement in the simulations by adjusting a single parameter compared to the simulations shown in Fig. 2: the angle of the magnetic tip field with the external field. While we found a very small angle of 8° for the tip used in Fig. 2, we find an angle of ~80° for the tip used in Extended Data Fig. 4. This means that when we approach the atom with the latter tip, we effectively add more in-plane field than out-of-plane field. Combined with the anisotropy of the hyperfine tensor, this results in qualitatively different behavior of the resonances with increasing tip field: the splitting between ESR resonances becomes smaller (comparable to Extended Data Fig. 2) and the NMR resonances, which are a direct measurement of this energy difference, shift to lower frequency.

**Data availability**
All data presented in this paper are publicly available at https://doi.org/10.5281/zenodo.8316339.

**Acknowledgements**
This work was supported by the Dutch Research Council (NWO Vici grant VI.C.182.016). P.W. acknowledges funding from the Emmy Noether Programme of the DFG (WI5486/1-1) and the Daimler and Benz Foundation.

**Author contributions**
LMV, LF and SO conceived the experiment. LMV, MPC and EWS performed the measurements. LMV, MPC and RB performed the calculations. All authors analyzed and discussed the results. LMV, LF, PW and SO wrote the manuscript, with input from all authors. SO supervised the project.

**Competing interests**
The authors declare no competing interests.

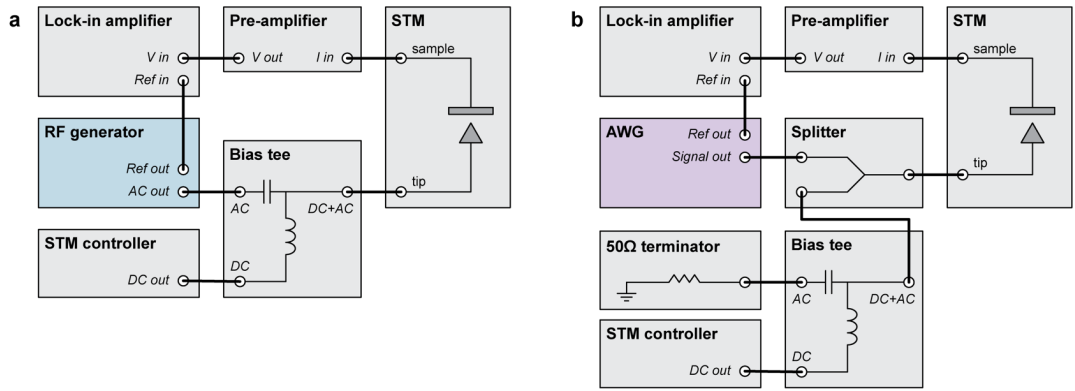

**Extended Data Fig. 1 | Wiring diagrams. a**, Schematic of electronics configuration during experiments in ESR-STM mode. **b**, Schematic of electronics configuration during experiments in pump-probe mode.

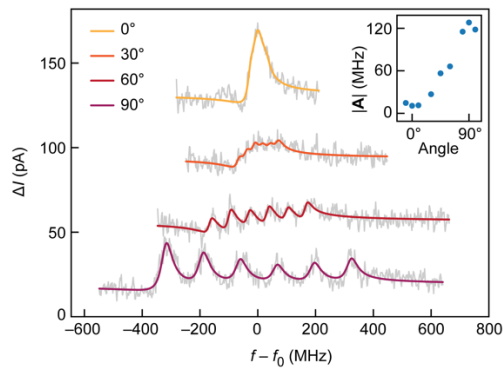

**Extended Data Fig. 2 | Hyperfine anisotropy measurements.** ESR-STM measurements on $^{47}$Ti at angles varying from in-plane (0°) to out-of-plane (90°) with fits of 6 equidistant Fano functions. Traces are offset for clarity. Inset: magnitude of hyperfine splitting obtained from fitting the entire dataset. All fitting parameters, except the hyperfine splitting magnitude and peak amplitude, were determined by first fitting an identical data set taken on a Ti isotope without nuclear spin. Experimental parameters: $I_{set}$ = 3 pA, $V_{set}$ = 60 mV, $V_{RF}$ = 25 mV, $T$ = 1.5 K, $|B_{ext}|$ = 0.5 –1.5 T.

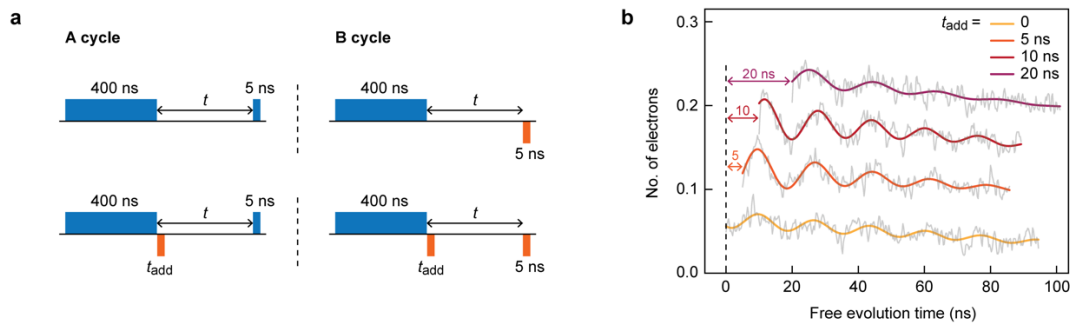

**Extended Data Fig. 3 | Pump-Probe pulse scheme. a**, Pulse schemes in the lock-in A and B cycles, without (top) and with (bottom) additional pump pulse. **b**, Comparison between measurements obtained using different widths $t_{add}$ for the additional pump pulse ($I_{set}$ = 4 pA, $V_{set}$ = 130 mV). Curves were shifted vertically for clarity. The free evolution time on the horizontal axis corresponds to the time $t$ indicated in (a). Measurements shown in the main paper were all performed using $t_{add}$ = 5 ns.

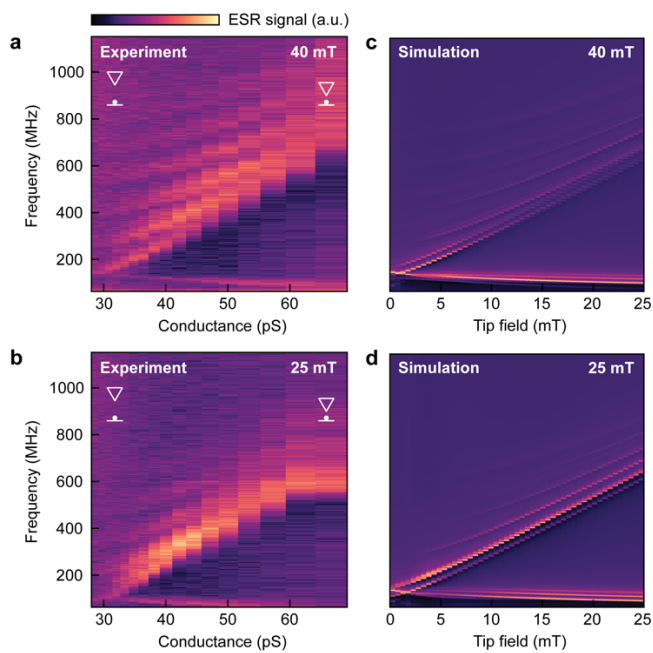

**Extended Data Fig. 4 | Different microtip. a**, ESR-STM measurements acquired on the same $^{47}$Ti atom as Fig. 2, but with a different microtip. Data taken at $B_{ext}$ = 40 mT, $T$ = 400 mK, $V_{RF}$ = 40 mV, $I_{set}$ = 2 pA. **b**, Same as (a), but taken at $B_{ext}$ = 25 mT. **c**, Simulation corresponding to (a). **d**, Simulation corresponding to (b). For the simulations we consider the tip field to make an 80° angle with the out-of-plane external field.